\documentstyle[11pt]{article}
\newcommand{\al}{\alpha}

\newcommand{\ep}{\epsilon}

\newcommand{\th}{\theta}

\newcommand{\la}{\lambda}

\newcommand{\De}{\Delta}

\newcommand{\La}{\Lambda}

\newcommand{\be}{\begin{eqnarray}}
\newcommand{\ee}{\end{eqnarray}}

\newcommand{\ti}{\tilde}

\newcommand{\np}{\newpage}
\newcommand{\hs}{\hspace}
\newcommand{\vs}{\vspace}

\newcommand{\nn}{\nonumber}

\newcommand{\hg}{\hat{g}}

\newcommand{\hpg}{\hat{g}^\prime}

\newcommand{\pri}{^\prime}
\newcommand{\lap}{\la^\prime}
\newcommand{\rhop}{\rho^\prime}

\newcommand{\Dg}{\Delta_g^+}
\newcommand{\Pro}{\prod_{n=1}^\infty (1-q^n)}
\newcommand{\Pg}{P^+_{\hg}}
\newcommand{\Pgp}{P^+_{\hg\pri}}
\newcommand{\hmu}{\hat{\mu}}
\newcommand{\hnu}{\hat{\nu}}
\newcommand{\hrho}{\hat{\rho}}
\newcommand{\gp}{g^\prime}
\newcommand{\pp}{\prime\prime}

\newcommand{\pa}{_\parallel}
\newcommand{\pe}{_\perp}
\begin{document}


\thispagestyle{empty}

\begin{flushright}
G\"oteborg ITP 94-8 \\
July 1994 \\
hep-th/9408087
\vs{10mm}\\
\end{flushright}

\begin{center}

{\huge{General branching functions of\vs{2mm} \\affine Lie algebras.}} \\
\vspace{10 mm}
{\Large{Stephen Hwang}\footnote{tfesh@fy.chalmers.se}} \vspace{2mm}\\
{\Large{and}}
\vspace{2mm}\\
{\Large{Henric Rhedin}\footnote{hr@fy.chalmers.se}} \\
\vspace{4 mm}

Institute of Theoretical Physics \\
Chalmers University of Technology \\
and \\
G\"oteborg University \\

\vs{15mm}

{\bf{Abstract}} \end{center}
\begin{quotation}
Explicit expressions are presented for general branching
functions for cosets of affine Lie algebras $\hat{g}$
with respect to subalgebras $\hat{g}^\prime$
for the cases where the corresponding finite
dimensional algebras $g$ and $g^\prime$ are such
that $g$ is simple and $g^\prime$ is either simple or
sums of $u(1)$ terms. A special case of
the latter yields the string functions. Our derivation
is purely algebraical and has its origin in the results on the
BRST cohomology presented by us earlier. We will here
give an independent and simple proof of the
validity our results. The method presented here generalizes in a
straightforward way to more complicated $g$ and $g^\prime$ such as
{\it e g } sums of simple and $u(1)$ terms.
\end{quotation}

\np

\setcounter{page}{1}
Since the discovery of Kac-Moody algebras \cite{Kac},\cite{Moody}
and the
 centrally
extended affine ones,
known as affine Lie algebras
(of which a first example was given in \cite{BH}), affine Lie
algebras
have played
an increasingly  important r\^ole in physics.
In the framework of conformal field theories, affine Lie algebras
appear
in the
study of string theory as well as critical phenomena in solid state
theory.
In particular, the so-called coset construction is crucial for the
description of known conformal field theories using affine Lie
algebras.
Examples
 of this
construction first appeared in \cite{BH} although a general form of
the stress-energy tensor was first given by \cite{GKO1}. It is in
this connection important to find the partition function
describing the
different cosets. In string theory they are nothing but the
zero-point
one-loop
amplitudes.
A slightly more general problem is to find the complete {\it
branching
functions}
of the different cosets, since from the knowledge of the latter
the partition
functions
are easily derived.

The purpose of this note is to present general expressions of the
branching
functions of $\hg$ with respect to a subalgebra $\hpg$ in the cases
when the
finite
dimensional algebras $g$ and $\gp$ are such that $g$ is simple
and $\gp$

is either
simple or a sum of $u(1)$ terms. In the latter case we can
specialize to the
Cartan subalgebra of $g$. Then the branching functions are
essentially the
{\it string functions}. The technique
which we will use here was first given in  \cite{HR} and the
present note
is a
rather straightforward application of this. However, we feel
that the
importance
of the branching and string functions motivates a derivation of
the explicit form for these. We will here also present an
alternative
and very
direct proof of the correctness of the  general expression.
The proof in
\cite{HR} is based on the computation of the  cohomology of
the BRST
operator
and, although correct, it is not very transparent for the present
application.

The explicit expressions for the branching and string functions
presented here
are
restricted solely to
the cases of {\it integrable representations}. The basic tool
which we
use apart
from
the construction in \cite{HR}
is the Weyl-Kac formula \cite{VK} of the character of an integrable
representation of an
affine Lie algebra. Expressions of branching and string functions in
special
cases have
appeared previously in the literature in particular in \cite{KP} and
\cite{KW}
(most of the results are reviewed in \cite{KII}). The general form of the
string function for the case of $\hat{su}(2)_k$ was first derived in
\cite{KP}.
In ref.\cite{BMP1} expressions for the
string functions of simply laced algebras as well as branching functions
for
several coset constructions was presented. The basic assumption in this
work is
the existence of a free field resolution of the irreducible affine
module. Such
a resolution has only been proven to exist for the case of $\hat{su}(2)$
\cite{BF}. The expressions
derived here for the string functions will confirm the results given in
 \cite{BMP1}. In ref.
\cite{KI} a general formula for
branching functions in terms of string functions was derived
\footnote{We thank
V.G. Kac for this reference.}. Using the explicit formulas for the
branching
as well as the string functions given here, one may verify these
relations. We will,
however, use our methods to give a simple derivation of the
branching functions directly in terms of string functions. These
expressions do not directly coincide with ref.\cite{KI},
but we expect that some
straightforward algebraic rearrangements will make them equal.
One may consider more general coset constructions than the ones
discussed here.
The
cases where instead of a simple subalgebra $g^\prime$ one has several
simple
terms $g^{\prime(1)}$, $g^{\prime(2)},\ldots$, yield branching functions
which are
easily derived using the methods presented here. This is true also in
the case
where
one or more of the simple terms are replaced with $u(1)$ terms. By
taking sums
of several simple algebras $g^{(i)}$, one may have many different cases.
In
\cite{R} a number of such cases will be presented: $\hg=\hg_{k_1}\oplus
\hg_{k_2}$
and  $\hpg=\hg_{k_1+k_2}$, $\hg$ and $\hg^\prime$ with
rank$g$=rank$g^\prime$,
$\hg=\left(\oplus_{a=1}^n \hg_{k_a}\right)$ and
$\hpg=\hg_{(\sum_{i=1}^nk_i)}$,
$\hg=\hg_{k_1}\oplus\hg^{\pp}_{k_2}$ and $\hpg=\hg^{\pp}_{k_1+k_2}$,
where
$g^{\pp}\subseteq g$. \vs{5mm}

Let $g$ be a simple finite dimensional Lie algebra and $\gp$ be a
subalgebra
of $g$. Their ranks are $r$ and $r\pri$. We denote by
$\hg$  and $\hg\pri$ the corresponding
affine Lie algebras. The levels $k$ and $k\pri$ of
$\hat{g}$ and $\hat{g}\pri$ are taken to be non-negative integers
($k=k^\prime$
for $\gp$ simple and $g$
simply laced). The set of roots of the finite algebra $g$ are
$\alpha\in \Delta_g$ and $\al\pri\in \Delta_{g\pri}$ for $\gp$. The
corresponding
 affine roots are $\hat{\alpha}\in\Delta_{\hg}$ etc. The
highest root of $g$ is denoted by $\psi_g$ and its length is taken
to be one.
The restrictions to positive roots are denoted $\Delta^+_g$ and
$\Delta^+_{g\pri}$. The number of elements in these sets are
$|\Delta^+_g|$,
$|\Delta^+_{g\pri}|$, respectively. We also define
$\Delta_{g,\gp}^+=\{\al\ |\ \al\in\Delta_g^+,\ \al\not\in
\Delta_{\gp}^+\}$ with the number of elements $|\Delta^+_{g,g\pri}|=
|\Delta^+_g|-|\Delta^+_{g\pri}|$.  The weight and root lattices of
$g$ and $\gp$
are denoted by $\Gamma_{w,g},\Gamma_{r,g},\Gamma_{w,\gp}$ and
$\Gamma_{r,\gp}$ etc
for $\hg$ and $\hat{g}\pri$. Define $\rho\in\Gamma_{r,g}$ by
$\{\rho\ |\
\al_i\cdot\rho=\al_i^2, \ \mbox{ for simple roots } \al_i\in
\Delta_g \}$.
$\rhop$,
$\hat{\rho}$ and $\hat{\rho}^\prime$ are the corresponding vectors
for
$\gp$, $\hg$ and $\hat{g}^\prime$. For the finite dimensional
algebras $\rho$ is
just the sum of positive roots.

Let $P^+_{\hg}$ and $P^+_{\hg\pri}$ be the set of
integrable highest weight representations of $\hg$ and $\hpg$ with
respective
highest weights $\hat{\la}\in\Gamma_{w,\hg}$ and
$\hat{\la}\pri\in\Gamma_{w,\hat{g}^\prime}$. Then
$P^+_{\hg}=\{\hat{\lambda}\mid \hat{\alpha}_i\cdot\hat{\lambda}\geq 0
\mbox{ for
simple roots } \hat{\alpha}_i\in \Delta_{\hg}\} $, or equivalently
$P^+_{\hg}=\{\lambda\mid \alpha_i\cdot\lambda\geq 0 \mbox{ for
simple roots }
\alpha_i\in \Delta_g \mbox{ and } \psi_g\cdot\lambda\leq k/2\}$. Let
$e_{\hat{\al}}$,  $f_{\hat{\al}}$ and $\hat{h}_i$, $\hat{\al}\in
\Delta_{\hg}^+$
and $i=0,1,\ldots,r$, be a realization of
$\hg$ according to the triangular decomposition
$\hg=\hat{n}_+\oplus\hat{n}_-  \oplus \hat{h}$ and $e_{\hat{\al}\pri}$,
$f_{\hat{\al}\pri}$ and $\hat{h}_{i\pri}$, $i\pri=0,1,\ldots,r\pri$,
be a
corresponding realization of $\hpg$. $\hat{h}_i,\ i=1,\ldots,r$ and
$\hat{h}_i^\prime,\
i=1,\ldots,r^\prime$ provide realizations of the Cartan
subalgebras $h$ and $h^\prime$ of $g$ and $\gp$. It is assumed
throughout for
all $\gp$ that $h\pri\subseteq h$. We will use the notation $\nu\pa$
for
the components of $\nu$ in $h\pri$ and $\nu\pe$ for the remaining
components.
Correspondingly, we will decompose the scalar product
$\nu\cdot\th=(\nu\cdot\th)\pa+(\nu\cdot\th)\pe$.

Let $L^g(\hat{\lambda})$ be an irreducible $\hg$-module of
highest weight $\hat{\lambda}$.
We define the character of $L^g(\hat{\lambda})$ \footnote{We are
omitting the
conventional factor $q^{-c/24}$ throughout this work.}
\be
\chi_\lambda^g(q,\theta)=\mbox{Tr}_{L^g(\hat{\lambda})}\left(q^{L_0}
e^{i\theta\cdot h} \right).
\ee
Here the trace is taken over all the vectors in $L^g(\hat{\lambda})$
and $L_0$
is the
zero mode of the Virasoro-generators of $\hg$ in the Sugawara
construction. In
the
case where $\hat{\lambda}\in P^+_{\hg}$ the character is given by
the Weyl-Kac
formula \cite{VK},
\be
\chi_\lambda^g(q,\th)&=&\sum_{t\in \La ^{\vee}_r}q^{\frac{(\la+
\rho/2+t)^2-
\rho^2/4}{k+c_g}}\sum_{w\in W_g}\ep(w)
e^{i(w(\la+\rho/2+t)-\rho/2)\cdot\th}R_g^{-1}(q,\th) \label{weylkac},
\ee
where $\La ^{\vee}_{r,g}$ denotes a lattice which is
spanned by $t=n(k+c_g)\al/\al^2$, where $\al\in\Delta_g^+$,
$n\in{\cal Z}$, and $c_g$ is the second Casimir of the adjoint
representation of $g$ ($c^\prime_g$ will denote the value for $\gp$).
$w$ are elements of the Weyl group $W_g$ and
\be
R_g(q,\th)=\prod_{n=1}^{\infty}(1-q^n)^{r}\prod_{\al\in\De_g^+}(1-
q^ne^{i\al\cdot\th})(1-q^{n-1}e^{-i\al\cdot\th}).
\ee
Define now the {\it coset module} $L^{g,\gp}(\hat{\lambda},
\hat{\lambda}\pri)$ by
\be
L^{g,g\pri}(\hat{\lambda},\hat{\lambda}\pri)=\{v\in L^g(\hat{\lambda})
\mid
e_{\hat{\al}\pri}(v)=0 \mbox{ and }
\hat{h}_{i}\pri(v)= \hat{\lambda}_i\pri v \mbox{ for
}\hat{\al}\pri\in\Delta_{\hat{g}\pri}^+\mbox{ and }  \hat{h}_i\pri\in
\hat{h}
\pri\}.
\ee
The branching function of this coset module is then defined as:
\be
b_{\lambda,\lambda\pri}^{g,g\pri}(q,\theta\pe)=\mbox{Tr}_{L^{g,g\pri}
(\hat{\lambda},\hat{\lambda}\pri)}
\left(q^{L_0^{g,g\pri}}e^{i(\theta\cdot h)\pe} \right).
\label{defbranch}\ee
We have here introduced the notation $L_0^{g,g\pri}= L_0^g-L_0^{g\pri}$.
The definition of the branching function implies, in the case of
integrable
representations, that the character of $L^g(\hat{\lambda})$
decomposes as
\be
\chi_{\lambda}^g(q,\theta)=\sum_{\la\pri\in
P^+_{\hat{g}\pri}}b_{\lambda,\lambda\pri}^{g,g\pri}(q,\theta\pe)\chi^
{g\pri}_{\la\pri}(q,\th\pa). \label{branch}
\ee
The main fact needed to show that eq.(\ref{defbranch}) implies eq.
(\ref{branch})
is that $L^g(\hat{\lambda})$ is isomorphic as a $\hg\pri$-module
to a direct sum
of modules
$L^{g\pri}(\hat{\lambda}\pri)$ with $\hat{\lambda}\pri\in P^+_{\hg\pri}$
(see
\cite{KP}, \S
4.9).

In \cite{HR} it was proven that the coset module $L^{g,g\pri}$ is
isomorphic to
the cohomology group ker$\hat{Q}$/Im$\hat{Q}$ of the so-called
relative BRST
operator $\hat{Q}$ acting on the module $M^{g}(\hat{\lambda})\otimes
M^{\tilde{g}}(\hat{\tilde{\lambda\pri}})\otimes M^{gh}$.
Here $M^{g}(\hat{\lambda})$ is the $\hg$ highest weight Verma module
and
$M^{\tilde{g}}
(\hat{\tilde{\lambda\pri}})$ is a $\hat{\tilde{g}}$ Verma module,
where
$\tilde{g}=\gp$ and   $\tilde{k}\pri=-k\pri-2c_{g}\pri$
and finally $M^{gh}$ is a module originating from the Faddeev-Popov
ghosts, which have been introduced in the BRST approach. The
isomorphism holds under the assumption that $\hat{\lambda}\in
P^+_{\hg}$ and
$-\hat{\tilde{\lambda\pri}}-\hat{\rho}\pri\in P^+_{\hat{g}\pri}$.
We will not
explain the details of the construction and proof of the isomorphism
here, but
refer to \cite{HR}. One should note, however, that the restriction
on the highest
weights $\hat{\tilde{\lambda\pri}}$ implies that
$M^{\tilde{g}}(\hat{\tilde{\lambda\pri}})$
 is irreducible, as can be seen from the Kac-Kazhdan determinant
\cite {KK}.
Instead we proceed more directly by giving the implications of the
isomorphism for the branching functions. First we define the following
characters
\be
\tilde{\chi}^{g\pri}_{\tilde{\lambda}\pri}(q,\theta\pa)&\equiv&
e^{i\tilde{\lambda}\pri\cdot\theta\pa}q^{-\frac{\tilde{\lambda}\pri\cdot
(\tilde{\lambda}\pri+\rho\pri)}{k\pri+c_{g}\pri}
}R^{-1}_{g\pri}(q,\theta\pa),\hs{5mm}-\tilde{\lambda}\pri
-\rho\pri\in P^+_{\hg\pri}\nonumber\\
\chi^{gh}(q,\theta\pa)&\equiv& e^{i\rho\pri\cdot\theta\pa}R_{g\pri}^2
(q,\theta\pa)\label{ghchar}\ee
for $g\pri$ simple and
\be
\tilde{\chi}^{g\pri}_{\tilde{\la}\pri}(q,\theta\pa)&\equiv&
e^{i\tilde{\la}\pri\cdot\theta\pa}q^{-\sum_{i=1}^{r\pri}\frac{
\tilde{\la}_i^
{\prime 2}}{k_i\pri}}
{1\over \Pro^{r\pri}}\nonumber\\
\chi^{gh}(q,\theta\pa)&\equiv& \Pro^{2r\pri}\ee
for $\gp=u(1)\oplus u(1)\oplus\ldots$ ($r\pri$ terms). Here
$k_i\pri$ are the
levels of $\hat{su}(2)$ such that the $u(1)$ terms are embedded
as Cartan
elements in the corresponding finite dimensional algebra
($k_i\pri=k$ for $g$
simply laced).
The characters are nothing but the characters of
$M^{g\pri}(\tilde{\lambda}\pri)$ and  $M^{gh}$ as discussed in
\cite{HR}. We introduce also a formal integration of exponents
defined
by\footnote{One may explicitely represent this formal integration
as the limit
$L\rightarrow\infty$ of ${1\over L^{r\pri}}\int_{-L/2}^{L/2}
d^{r^\prime}\theta
\pa.$} $\int d^{r\pri}\th\pa \hs{1mm}e^{i\theta\pa\cdot \nu}
\equiv\prod_{i=1}
^{r\pri}\delta_{\nu_i,0}\equiv\delta_{\nu,0}.$\vs{5mm}\\
{\sc Proposition 1. } {\it With the notation above and $\gp$
simple or $\gp=u(1)\oplus u(1)\oplus u(1)\ldots $ ($r\pri$ terms)
we have} \be
b_{\lambda,\lambda\pri}^{g,g\pri}(q,\th\pe)=\int d^{r\pri}\th\pa
\chi^g_{\lambda}
(q,\th)
\ti{\chi}^{g\pri}_{\ti{\lambda}\pri}(q,\th\pa)\chi^{gh}(q,\th\pa)
\label{brstchar}
\ee
{\it where $\hat{\la}\pri\equiv
-\hat{\tilde{\la\pri}}-\hat{\rho}\pri\in\Pgp,\ \hat{\la}\in\Pg$.}
\vs{5mm}\\
{\it Proof.} Using eq.(\ref{branch}) the right-hand side of eq.
(\ref{brstchar})
reads
\be
\sum_{\la\pri\in\Pgp}b_{\lambda,\lambda\pri}^{g,g\pri}(q,\theta\pe)
\int
d^{r\pri}\th\pa
\chi^{g\pri}_
{\la\pri}(q,\th\pa)
\tilde{\chi}^{g\pri}_{\tilde{\lambda}\pri}(q,\theta\pa)\chi^{gh}(q,
\theta\pa).
\ee
since the branching function is independent of
the variables integrated over. For $\gp$ simple we use the Weyl-Kac
formula
eq.(\ref{weylkac}) for $\chi^{g\pri}_{\la\pri}$ and eq.(\ref{ghchar})
to perform
the integration. The right-hand side of eq.(\ref{brstchar}) is then
$$
\sum_{\la\pri\in\Pgp}\hs{-1mm}b_{\lambda,\lambda\pri}
^{g,g\pri}(q,\th\pe)\hs{-2mm}\sum_{t\pri\in\La_{r,g\pri}^{\vee}}
\sum_{w\in
W_{g\pri}}\hs{-2mm}\epsilon(w)
q^{{(\la\pri+\rho\pri/2+t\pri)^2-\rho^{\prime 2}/4\over k\pri+
c\pri_g}}
q^{-\frac{\tilde{\lambda}\pri\cdot
(\tilde{\lambda}\pri+\rho\pri)}{k\pri+c_{g}\pri}} \delta_{[w(\la
\pri+\rho\pri/2+
t\pri)-\rho\pri/2+\ti{\la}\pri+\rho\pri],0}.
$$
For $\hat{\mu}\pri$,$\hat{\nu}\pri\in \Pgp$ and $w\in W_{\hpg}$,
the affine Weyl
group of $\gp$, we have that
the equation $w(\hmu\pri+\hrho\pri /2)-\hrho\pri /2=\hnu\pri$
implies that
$\hmu\pri=\hnu\pri$ and $w=Id.$ (for a proof
in the finite dimensional case see {\it e.g.} \cite{JH}, lemma A
in section 13.2
and for the affine case \cite{KII}, lemma 10.3 together with prop.
3.12b ). Then
$$\sum_{t\pri\in\Lambda_{r,g\pri}^\vee}\sum_{w\in
W_{g\pri}}\epsilon(w)
q^{{(\la\pri+\rho\pri/2+t\pri)^2-\rho^{\prime 2}/4\over k\pri+
c\pri_g}}q^{-
\frac{\tilde{\lambda}\pri\cdot
(\tilde{\lambda}\pri+\rho\pri)}{k\pri+c_{g}\pri}}\delta_{[w(
\lambda\pri+
\rho\pri/2+t\pri)-\rho\pri/2+\tilde{\lambda}^\prime+\rho^\prime],0}=
\delta_{\lambda^\prime+\tilde{\lambda}^\prime+\rho^\prime ,0}$$
which proves the
proposition for the case of $\gp$ being simple. The case $\gp=u(1)
\oplus u(1)
\oplus\ldots\oplus u(1)$ is completely
analagous.\ \ $\Box$\vs{5mm}\\
{\sc Theorem.} {\it Let $g$ be a simple
finite dimensional algebra, $\hg$ its affine extension,
$\hat{\lambda}\in\Pg$
the highest weight of a representation of $\hg$, $\hat{g}\pri$ a
subalgebra of
$\hg$ with $h\pri\subseteq h$ and $\hat{\la}\pri\in\Pgp$. Then
the affine
branching function
$b_{\lambda,\lambda^\prime}^{g,g^\prime} (q,\theta\pe)$ is
\vs{2mm}\\
(i) $\gp$ simple:
\be
b_{\lambda,\lambda\pri}^{g,g^\prime}(q,\theta\pe)=
{1\over\Pro ^{r-r\pri+2|\Delta^+_{g,\gp}|}}\sum_{t\in
\Lambda_{r,g}^{\vee}}
\sum_{w\in
W_g}
\hs{-2mm}\epsilon(w)q^{{(\lambda+\rho/2+t)^2-\rho^2/4\over
k+c_g}}q^{-{(\lap+\rhop/2)^2 -\rho^{\prime 2}/4\over k^\prime+
c_g^\prime}}
\nonumber\ee \be
\hs{0mm}e^{i\left((w(\lambda+\rho/2+t)-\rho/2)\cdot\theta\right)\pe}
\prod_{\alpha\in\Delta_{g,g^\prime}^+}\hs{-1mm}\sum_{p_\alpha}
\sum_{s_\alpha=0}^\infty
\left((-1)^{s_\alpha}q^{\frac{1}{2}[(s_\alpha-p_\alpha+1/2)^2-
(p_\al-1/2)^2]}
e^{ip_\alpha(\alpha
\cdot\theta)\pe}\right)\label{branchf}\ee where $p_\alpha\in
{\cal Z}$ is
restricted
by \be
\sum_{\alpha\in \Delta_{g,g^\prime}^+}p_\alpha\alpha\pa+
(w(\lambda+\rho/2+t)-\rho/2)\pa
-\lap=0\label{pcond}.\ee
\vs{2mm}\\
 (ii) $\gp=u(1)\oplus u(1)\oplus\ldots$ ($r\pri$ terms) :
\be
b_{\lambda,\la\pri}^{g,g^\prime}(q,\theta\pe)=
{1\over\Pro ^{r-r\pri+2|\Dg|}}\sum_{t\in\Lambda_{r,g}^{\vee}}
\sum_{w\in W_g}\epsilon(w)q^{{(\lambda+\rho/2+t)^2-\rho^2/4\over
k+c_g}-
\sum_{i=1}^{r\pri}{\la^{\prime 2}_i
\over k^\prime_i}}\nonumber\\
e^{i\left((w(\lambda+\rho/2+t)-\rho/2)\cdot\theta\right)\pe}
\prod_{\alpha\in\Delta_{g}^+}\sum_{p_\alpha}\sum_{s_\alpha=0}^
\infty
\left((-1)^{s_\alpha}q^{\frac{1}{2}[(s_\alpha-p_\alpha+1/2)^2-
(p_\al-1/2)^2]}
e^{ip_\alpha(\alpha
\cdot\theta)\pe}\right)\label{branu}\ee
where $p_\alpha\in {\cal Z}$ is restricted by
\be
\sum_{\alpha\in\Delta_{g}^+}p_\alpha\alpha_{\pa}+(w(\lambda+
\rho/2+t)-\rho/2)\pa
-\la\pri=0\ee \vs{5mm}\\
Proof:} We use eq.(\ref{brstchar}). In order to perform the
integrations we first write
\be
R_g(q,\th)=R_{\gp}(q,\th\pa)\Pro^{r-r^\prime}
\prod_{\alpha\in\Delta_{g,\gp}^+}(1-q^ne^{i\al\cdot\th})(1-q^{n-1}
e^{-i\al\cdot\th})
\ee
and use the identity \cite{CP}-\cite{BCMN}
\be
 \prod_{\alpha\in\Delta_{g,\gp}^+}(1-q^ne^{i\al\cdot\th})^{-1}
(1-q^{n-1}
e^{-i\al\cdot\th})^{-1}=\hs{55mm}\nn\\{1\over\Pro^{2|\Delta_{g,
\gp}^+|}}
\prod_{\alpha\in\Delta_{g,\gp}^+}\sum_{p_\alpha}
\sum_{s_\alpha=0}^\infty
(-1)^{s_\alpha}q^{\frac{1}{2}[(s_\alpha-p_\alpha+1/2)^2-(p_\al-
1/2)^2]}
e^{ip_\alpha\alpha\cdot\theta},\hs{0mm}\label{prodid}
\ee
where $p_{\alpha}\in{\cal Z}$ for every $\alpha\in\Delta_{g,\gp}^+
$, to rewrite
the
expression for the character of $L^g(\hat{\la})$ eq.(
\ref{weylkac}). Then this
together with eq.(\ref{ghchar}) inserted into eq.(\ref{brstchar})
yields for the case of $\gp$ simple
\be
b_{\lambda,\lambda^\prime}^{g,g^\prime}(q,\theta\pe)&=&
{1\over \Pro^{r-r\pri+2|\Delta^+_{g,\gp}|}}\hs{1mm}
\sum_{t\in\Lambda_{r,g}^{\vee}}q^{(\lambda+\rho/2+t)^2-\rho^2/4
\over k+c_g}
q^{-{(\tilde{\lambda}^\prime+\rhop/2)^2-\rho^{\prime 2}/4\over
k^\prime+c_g^\prime}}\nn\\
&&\sum_{w\in W_g}
\epsilon(w)\int d^{r^\prime}\theta\pa
e^{i(w(\lambda+\rho/2+t)-\rho/2)\cdot\theta}
e^{i(\tilde{\lap}+\rhop)\cdot\th\pa}\nn\\
&&\prod_{\alpha\in\Delta_{g,\gp}^+}\sum_{p_\alpha} \sum_{s_\alpha
=0}^\infty
(-1)^{s_\alpha}q^{\frac{1}{2}[(s_\alpha-p_\alpha+1/2)^2-(p_\al-
1/2)^2]}
e^{ip_\alpha\alpha\cdot\theta}.
\ee
Upon integrating over $\th\pa$ we will
get eq.(\ref{branchf}) and deltafunctions that impose the condition
eq.(\ref{pcond}), where $\la^\prime =-\tilde{\la}^\prime-\rhop$.
The proof of (ii) is completely analagous.\ \
$\Box$\\

The important special case of (ii) above is when the branching
functions are the string functions (up to a factor $\eta^{-r}$,
where $\eta$ is
the Dedekind function). The latter are then given by eq.(
\ref{branu}) with
$r\pri=r$ and read explicitely
\be
c_{\lambda,\la\pri}^{g}(q)=
{\eta^{-r}(q)\over\Pro ^{2|\Dg|}}\sum_{t\in\Lambda_{r,g}^{\vee}}
\sum_{w\in W_g}\epsilon(w)q^{{(\lambda+\rho/2+t)^2-\rho^2/4\over
k+c_g}-
\sum_{i=1}^r{\la^{\prime 2}_i
\over k^\prime_i}}\nonumber\\
\prod_{\alpha\in\Delta_{g}^+}\sum_{p_\alpha}\sum_{s_\alpha=0}^\infty
\left((-1)^{s_\alpha}q^{\frac{1}{2}[(s_\alpha-p_\alpha+1/2)^2-(p_
\al-1/2)^2]}
\right),\ee
where $p_\alpha\in {\cal Z}$ is restricted by
\be
\sum_{\alpha\in
\Delta_{g}^+}p_\alpha\alpha+w(\lambda+\rho/2+t)-\rho/2 -\la\pri=0.
\vs{5mm}\ee
Let us comment that there are completely analagous expressions
for the more general case of $\gp$ consisting of a mixture of
several simple or
$u(1)$
terms, but we have chosen not give the expressions explicitely.
They may
be derived in the same straightforward manner as above.

We finally derive an expression for the branching functions in
terms of the string
functions.\vs{5mm}\\
{\sc Proposition 2. } {\it With the same notation as in the
theorem, we have
the branching function for the case of $g\pri$ simple}
\be
b_{\lambda,\lambda\pri}^{g,g^\prime}(q,\theta\pe)=
{q^{\frac{r}{24}}\over\Pro ^{|\Delta^+_{g\pri}|-2r-r\pri}}\hs{1mm}
q^{-\frac{\la\pri(\la\pri+\rho\pri)}
{k\pri+c_g\pri}}\hs{38mm}
\nonumber\\
\sum_{\mu\in\Gamma_{w,g}}c^g_{\la,\mu}(q)\hs{1mm}q^{\sum_{i=1}^
r\hs{-1mm}
\frac{\mu_i^2}{k_i}}
e^{i(\mu\cdot\th)\pe}\prod_{\al\pri\in\Delta^+_{g\pri}}\sum_{p_
{\al\pri}}
(-1)^{p_{\al\pri}}q^{\frac{1}{2}[(p_{\al\pri}+1/2)^2-1/4]}\ee
{\it where $p_{\al\pri}\in{\cal Z}$ is restricted by}
\be
\sum_{\al\pri\in\Delta^+_{g\pri}}p_{\al\pri}\al\pri+\mu\pa-\la\pri=0.
\ee
\vs{5mm}
{\it Proof:} The proof of the proposition is completely analagous
to that
of the theorem. However, instead of the Weyl-Kac formula for
the character of $\hat{g}$ we take eq.(\ref{branch}) with $g\pri=h$
\be
\chi^g_\la (q,\th)=\sum_{\mu\in\Gamma_{w,g}}b^{g,h}_{\la,\mu}(q)e^
{i\mu\cdot\th}
q^{\sum_{i=1}^r\frac{\mu_i^2}{k_i}}\Pro^{-r}.
\ee
Furthermore, in order to perform the integration we use the
following
identity in place of eq.(\ref{prodid})
\be
R_{g\pri}(q,\th\pa)=\prod_{n=1}^\infty (1-q^n)^{r\pri-|\Delta_
{g\pri}^+|}
\prod_{\al\pri\in
\Delta_{g\pri}}\sum_{p_{\al\pri}\in {\cal Z}}(-1)^{p_{\al\pri}}
q^{\frac{1}{2}[(p_{\al\pri}+1/2)^2-1/4]}e^{ip_{\al\pri}\al\pri
\cdot\th\pa },\ee
which is easily derived using the Jacobi triple product identity
$$\sum_{p=-\infty}^\infty q^{p^2}e^{ip\th }=\prod_{p=1}^\infty
(1-q^{2p})
(1+q^{2p-1}e^{i\th})(1+q^{2p-1}e^{-i\th}).$$
\begin{flushright} $\Box$\end{flushright}

\np

\end{document}